\documentclass{article}
\usepackage{graphicx}
\usepackage[round]{natbib}
\usepackage{epsfig}
\title{Five year ahead prediction of Sea Surface Temperature in the Tropical
Atlantic: a comparison between IPCC climate models and simple
statistical methods}

\author{
Thomas Laepple (AWI)\\
Stephen Jewson (RMS)\footnote{\emph{Correspondence email}: \texttt{x@stephenjewson.com}}\\}

\begin{document}
\maketitle

\begin{abstract}
There is a clear positive correlation between boreal summer tropical Atlantic sea-surface
temperature and annual hurricane numbers. This motivates the idea of trying to predict
the sea-surface temperature in order to be able to predict future hurricane activity.
In previous work we have used simple statistical methods to make 5 year predictions
of tropical Atlantic sea surface temperatures for this purpose.
We now compare these statistical SST predictions with SST predictions
made by an ensemble mean of IPCC climate models.
\end{abstract}

\section{Introduction}

The insurance and reinsurance industries use predictions of annual numbers of landfalling hurricanes
to help them set the
rates on the contracts they sell, and to allocate sufficient
capital to their businesses.
This creates a need to produce better predictions of future hurricane numbers,
especially over the 0-5 year time-scales of interest to these industries.
We have been looking into a number of ways of making such predictions, based on several different
approaches such as time-series analysis historical landfalling hurricane data, time-series
analysis of basin hurricane number data, and statistical predictions of sea-surface temperature (SST).
In the latter approach, time-series methods are used to predict future SSTs based on historical SSTs,
and the predictions of SST are then converted to predictions of landfalling hurricane numbers
using statistical relations. One obvious question arises: might it not be better to use numerical
models of climate, rather than statistical methods, to predict future SSTs? We feel that the answer
to this question is not at all obvious, and the purpose of this article is to perform an initial
comparison of numerical model predictions with our simple statistical methods.
Our approach is to take simulations of the 20th century climate
from a set of integrations of state-of-the-art climate models
that were prepared as part of a submission to the
Intergovernmental Panel for Climate Change (IPCC: see~\texttt{www.ipcc.ch}),
and compare the predictions from these models
with those from our own simple statistical methods.

\section{Methods}

\subsection{Data}

We are trying to predict SSTs in the Main Development Region (MDR) for hurricanes,
defined as 10$^o$-20$^o$N, 15$^o$-70$^o$W.
From a statistical analysis of the correlation of SSTs in different months with observed hurricane numbers,
we choose an index based on the average SST in this region for the period July to September. This index
has one value per year, and extends from 1860 to 2005, although we only use values from 1900 to 2000.
We derive values for this index from the HADISST data set~\citep{hadisst}.

\subsection{Statistical models}

The statistical model we use for this comparison is taken from a comparison of the ability of simple
statistical models to predict this MDR SST index, described in~\citet{e20}. The winning model in this comparison
was the `8 year flat-line', which predicts future SSTs using a simple average of SSTs from the last 8
years. For a 5-year forecast this model beats models with longer and shorter averaging windows, and beats models that attempt
to model any trends in the data. Why does this model do so well? There is considerable interannual,
decadal and multidecadal time-scale variability in the SST index. Since we are trying to predict SSTs over the next 5
years, and the interannual variability is only predictable over periods of a few months, making a 5 year
prediction is mainly about estimating the current level of the decadal and interdecadal variability.
The 8 year window presumably works because it does this well. A shorter window would be too influenced
by interannual variability during the windowing period, and a longer window would presumably fail to capture
the current level of the long-term variability.

\subsection{Numerical models}

The numerical models we use for this comparison are coupled-ocean atmosphere models, running from effectively
random initial conditions in 1900 throughout the 20th century. The models are forced with estimates
of the observed climate forcings in place during the 20th century, the most important of which are
changing levels of CO$_2$, sulphate aerosols, solar variability and volcanic activity.

The methodology we use to make predictions from the climate model output is as follows:

\begin{itemize}

    \item For each of the 22 coupled climate models simulations of the 20th century available
    from the IPCC, we take the 101 year time-series of simulated MDR SST index values.

    \item We split the simulations into two sets: those that include volcanic forcing,
    and those that do not

    \item We create an ensemble mean of the climate model simulations in each set

    \item In order to create a prediction of real MDR SSTs from year $n+1$ onwards, we calculate
    a bias correction to the ensemble mean climate model predictions based on the 8 years $n-8, n-7, ...,n-1, n$

    \item Applying this bias correction, we then predict future MDR SST values from simulated values
    for the years $n+1$ onwards.

\end{itemize}

We calibrate using 8 years of data to make the comparison with the statistical model as clean as
possible.
As a result of this calibration, if the climate model predicted constant values it would give exactly the same prediction
as the statistical model. It will give better predictions if the fluctuations produced by the model
are, on average, realistic.

There are two reasons why this methodology might lead to skillful predictions.
First, the bias correction sets the level of any internal variability in the predictions.
For instance, if there were a multidecadal cycle in observed MDR SSTs, the bias correction
will put the predictions at the right level within that cycle. This is essentially the same
source of predictive skill as is exploited by the statistical model. Note, however, that
the climate model predictions would not be expected to continue the development of
such a cycle into the future in a realistic way, since they start from random initial conditions, and know nothing
about the phase of any cycles of internal climate variability.
Second, to the extent that there is a part of climate variability that is driven by variations
in the external forcings, then the models might be able to capture that. Most obviously, if the
rise in CO$_2$ during the 20th century drives any changes in MDR SSTs, then the models could in
principle reproduce such a change.

By using simulations of past climate, driven by after-the-fact estimates of the forcings,
we are effectively assuming that the forcings driving these models are
perfectly predictable, which gives these models a large unfair and unrealistic advantage over the statistical models.
The least predictable of the forcings is the volcanic forcing, which is in reality totally unpredictable.
This is why we split the climate models into two sets.
Comparing the statistical model with the set that doesn't include volcanoes
gives a much fairer comparison, since the remaining forcing parameters are \emph{reasonably} predictable.
It still, however, gives a slightly unfair advantage to the climate models.

\begin{figure}[!hb]
 \begin{center}
    \scalebox{0.7}{\includegraphics{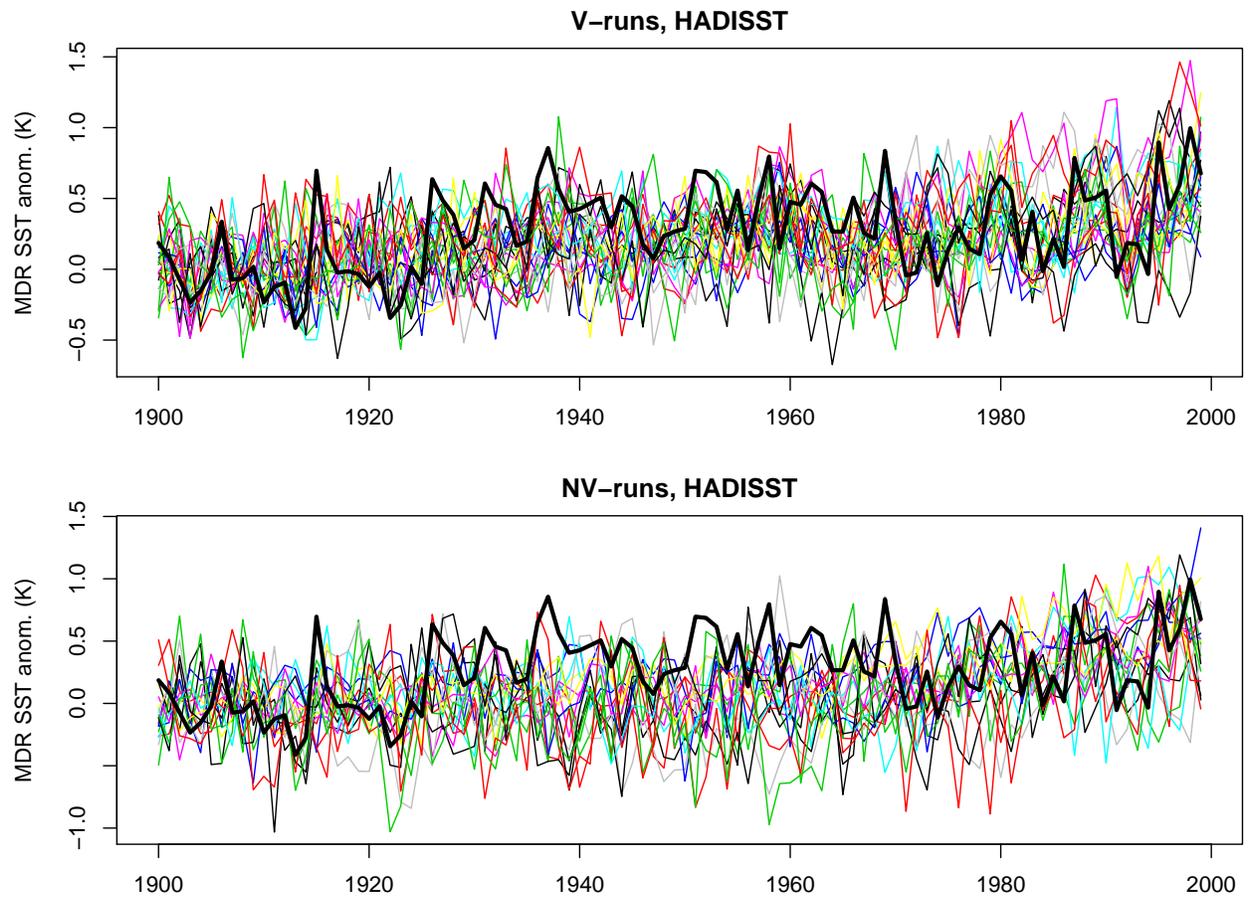}}
  \end{center}
    \caption{
Upper panel: MDR SST simulations from the `with volcanic forcing' IPCC climate model runs we use in our comparison, along
with observed MDR SSTs (solid line).
Lower panel: the same, but for `without volcanic forcing' runs.
}
     \label{f01}
\end{figure}

\begin{figure}[!hb]
 \begin{center}
    \scalebox{0.7}{\includegraphics{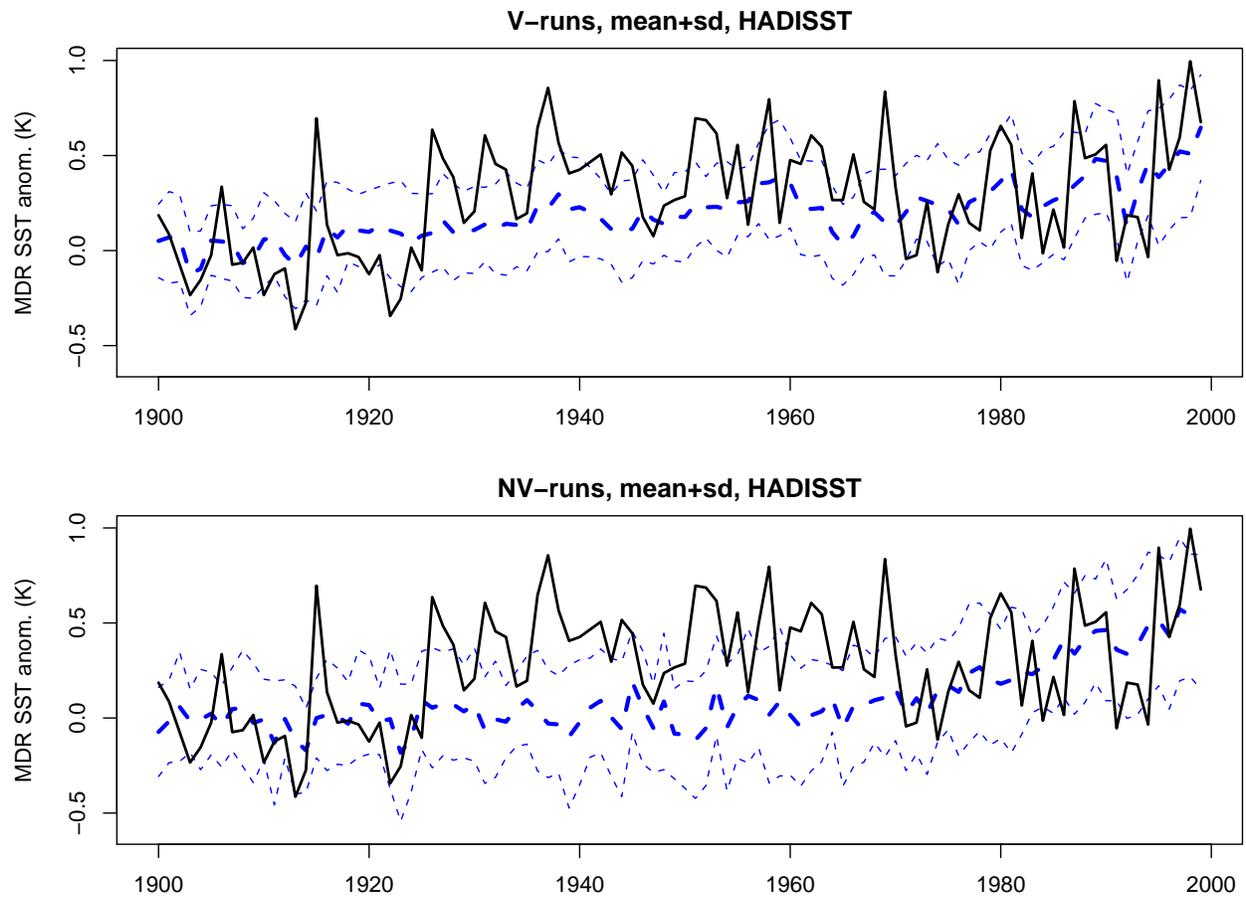}}
  \end{center}
    \caption{
As figure~\ref{f01}, but with the climate model simulations summarised as a mean and plus and minus one standard deviation.
}
     \label{f02}
\end{figure}

\begin{figure}[!hb]
 \begin{center}
    \scalebox{0.7}{\includegraphics{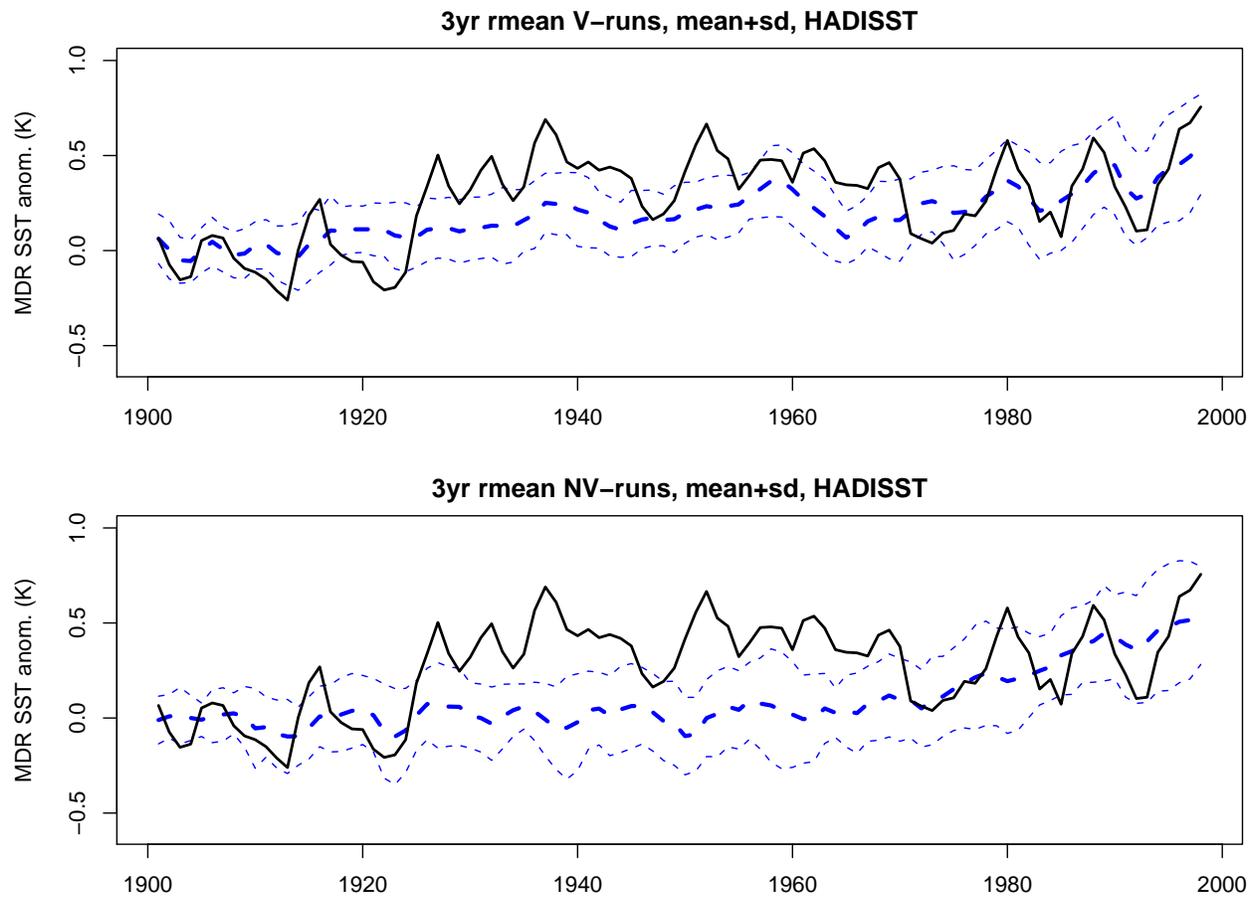}}
  \end{center}
    \caption{
As figure~\ref{f02}, but smoothed with a 3-year running mean.
}
     \label{f03}
\end{figure}

\begin{figure}[!hb]
 \begin{center}
    \scalebox{0.7}{\includegraphics{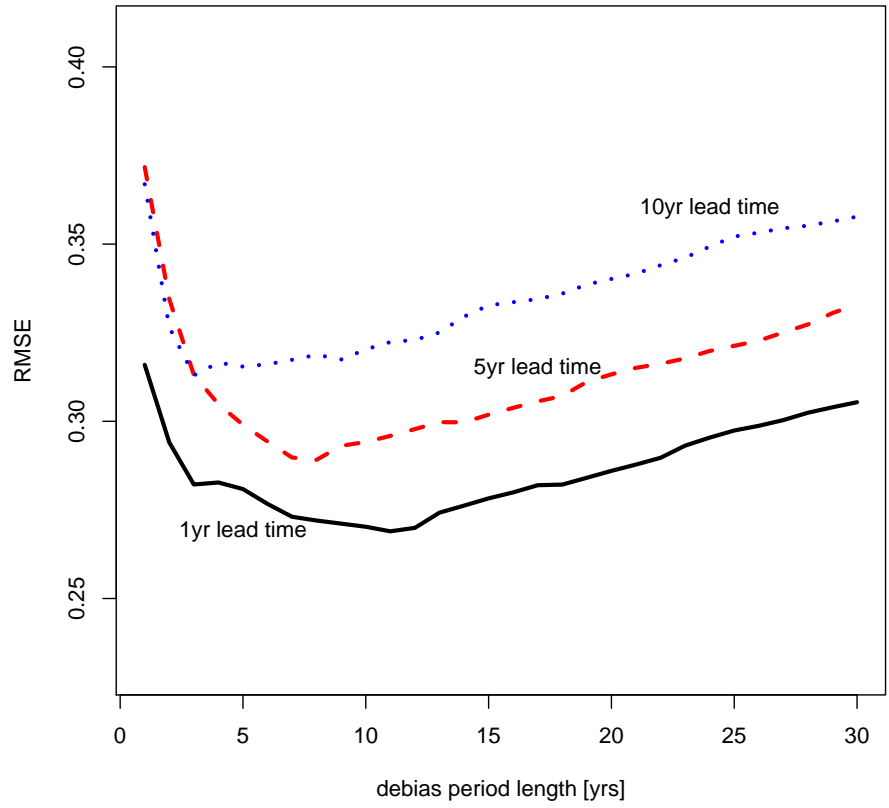}}
  \end{center}
    \caption{
Predictive performance of the ensemble mean of the `without volcanic forcing' climate model runs,
versus length of the calibration window.
}
     \label{f04}
\end{figure}

\begin{figure}[!hb]
 \begin{center}
    \scalebox{0.7}{\includegraphics{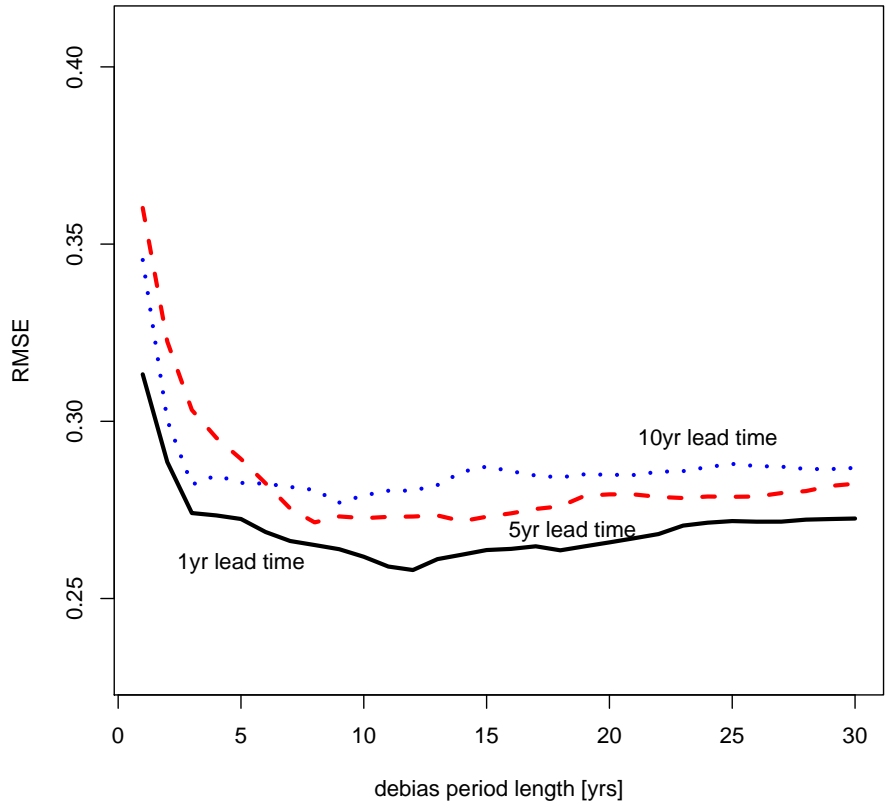}}
  \end{center}
    \caption{
As figure~\ref{f04}, but for the `with volcanic forcing' climate model runs.
}
     \label{f05}
\end{figure}

The climate model simulations of MDR SST from the runs with volcanic forcing are shown in
the top panel of figure~\ref{f01}, along with the observed MDR SST (solid black line).
The climate model simulations of MDR SST from the runs without volcanic forcing are shown in the
lower panel. Figure~\ref{f02} shows the same data, but with the range of results from individual climate
models shown as the mean plus and minus one standard deviation.
Figure~\ref{f03} shows the same data again, but smoothed with a 3 year running mean to emphasize longer timescales.
Figures~\ref{f04} and~\ref{f05} show the forecast errors for the climate models versus
the length of the calibration period. The optimal calibration period is longest
for the shortest lead times. The 8 year calibration period used for our comparison
is apparently optimal for forecasts with lead times of around 5 years, and seems to be a good
overall choice.

Based on the results in figures~\ref{f04} and~\ref{f05} one could
consider varying the length of the calibration window for the climate model according to the lead-time
being predicted. One would also then have to vary the length of the window in the statistical
method in a similar way, to keep the comparison fair.
We, however, consider this unnecessarily complex at this point.

\section{Results}

\begin{figure}[!hb]
 \begin{center}
    \scalebox{0.7}{\includegraphics{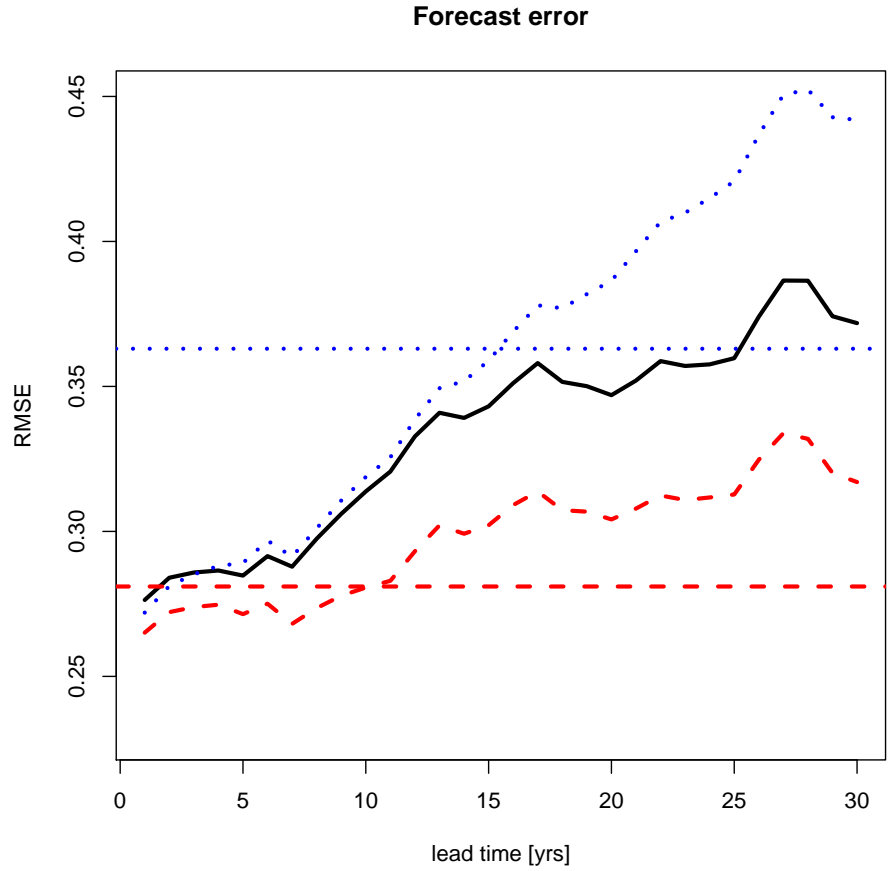}}
  \end{center}
    \caption{
Predictive performance of the statistical model (solid black line), climate model
ensemble mean with volcanic forcing (dotted blue line) and climate model ensemble mean
without volcanic forcing (dashed red line).
The horizontal lines show the performance of the climate model predictions when
calibrated using the entire data-set (included for reference only since this is not a proper
out-of-sample calibration method).
}
     \label{f06}
\end{figure}

The results of our comparison between the predictive skill of our simple statistical method and the climate models
are given in figure~\ref{f06}. The solid black line shows the RMSE of the forecasts from our
statistical model versus lead time.
The error in these forecasts gradually increases with lead time, although
the error for 30 year predictions is only around 30\% larger than that for 1 year predictions.
The extra skill at one year is presumably because the model is capturing some of the
decadal and interdecadal time-scale variability in the time-series, including any long-term trends.
The error at one year is presumably dominated by the interannual variability in the climate, which
is not predicted by this model.

The dashed red line then shows the RMSE of the forecasts derived from the climate models that include
a perfect prediction of future volcanic activity. These models do very slightly better than the
statistical models at all lead times: the errors are around 5\% smaller.

The dotted blue line shows the RMSE of the forecasts derived from the climate models that do not
include volcanic activity, but do include a perfect prediction of the other atmospheric forcings.
In practice, it is not possible to predict any of the atmospheric forcings perfectly, and these
RMSE values should thus be considered as artificially too low, especially at longer lead-times.
We see that the climate models do very slightly better than the statistical model up to around 5 years,
and are worse than the statistical model beyond 5 years.

Figure~\ref{f06} shows the same results, but now with error bars. We see that at short lead-times
the differences between the different models are not significant.

What can we understand from these results, and in particular the result that the climate models without
volcanic forcing don't perform materially better than the statistical model?
The most obvious interpretation is that the non-volcanic climate model simulations contain
no information about the direction that climate is moving.
Their skill comes purely from the bias correction, that sets them
at the correct current level for climate.
The fact they do progressively \emph{worse} than the statistical model as the lead time increases is
presumably because of gradual drift in the models.

Why do the climate models do so badly? The two extreme-case explanations are
(a) that the models are perfect representations of the physics of the real climate,
but that climate variability is driven by internal climate variability, rather than externally forced
variability, and
(b) that the models are very poor representations of the physics of the real climate,
and even if there is some part of climate variability that is driven by external forcings,
the models fail to capture it.
Reality is presumably a combination of these two.

The volcanically forced models, on the other hand, do have some information about which way the climate moves
away from the 8 year baseline. It is clear that these models do have some (albeit retrospective)
skill in capturing the climate response
to volcanoes.

\section{Conclusions}

We are developing methods to predict hurricane numbers over the next 5 years.
One class of methods we are developing is based on the idea of predicting main development region SSTs
over this time period, and then converting those predictions to predictions of hurricane numbers.
Hitherto we have used simple statistical methods to make our SST predictions. In this article, however,
we have considered the possibility of using climate models instead. We have compared predictions
made from an ensemble of state-of-the-art climate models with our statistical
predictions. Climate models that include a perfect prediction of future volcanic activity are the best
predictions of those that we consider. Climate models that ignore volcanic activity, but include a perfect
prediction of other climate forcings, do very slightly better than our statistical models up to a lead time of
5 years, but do increasing less well thereafter. Given the unfair advantage that using perfect predictions
of future atmospheric forcing parameters confers on the climate models, the tiny margin by which they
beat the statistical models, and the vast complexity of using these models versus using the statistical models,
we conclude that the statistical models are preferable to the climate models over our time period of interest.

We have thus answered our initial question. Should we use IPCC climate model output to feed into our
predictions of future hurricane numbers, rather than the simple statistical methods we currently use? The
answer is no.

What can we say about (a) whether the climate models might one day be preferable to the statistical models
and (b) what this tells us about predictions of climate in the 21st century?
Wrt (a), the one obvious way that the climate model predictions \emph{might} be improved would be to include realistic,
rather than random, initial conditions. If there really is a component of unforced
internal climate variability that is predictable on decadal and interdecadal time-scales, and the models
can simulate it, then their errors might be reduced. However, whether predictable internal climate
variability exists on these long-timescales is currently unknown, and climate scientists have different
opinions on this subject.
In the absence of such signals, at short lead-times both the statistical models and the climate models
are probably hitting the limits of predictability defined by the levels of interannual variability in the climate,
that is unpredictable over the time-scales of interest.
At long-time scales it seems likely that the climate models could be significantly improved,
by the gradual eradication of the slow drift that is presumably driving the more rapid increase in the climate
model errors versus the errors in the statistical model.

Wrt (b), should we extrapolate these results to tell us about the likely skill of future climate predictions?
This is rather hard to answer, since it depends, in part, on what the climate does in the future. But,
\emph{prima facie}, the suggestion from these results is that statistical models will continue to do as well or
better than climate models in the future.
However, what if future climate variability is dominated by single externally-forced upward trend, as many
climate models suggest,
rather than by interdecadal variability, as it has been in the past century?
Will the climate models then do any better? Perhaps. Although at that point
statistical models that capture trends would improve as well, and the comparison between statistical models
and climate models would have to be repeated.

\bibliography{arxiv}

\newpage

\newpage

\newpage

\newpage

\newpage

\newpage


\newpage
\begin{figure}[!hb]
 \begin{center}
    \scalebox{0.7}{\includegraphics{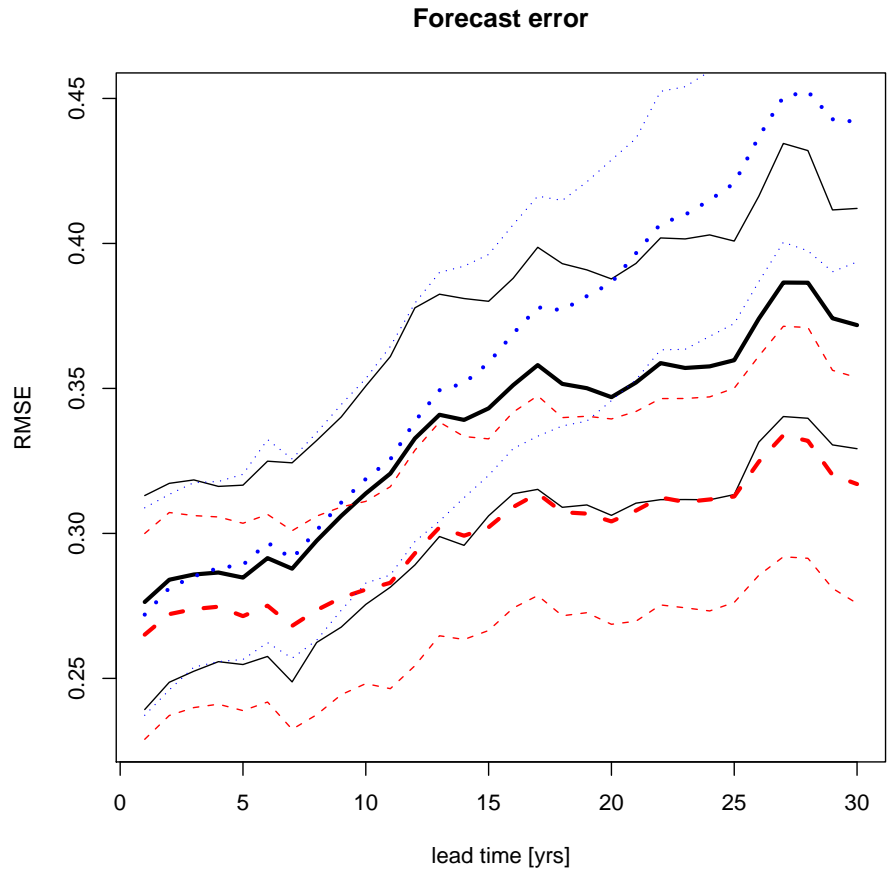}}
  \end{center}
    \caption{
As figure~\ref{f06}, but with error bars on each prediction.
}
     \label{f08}
\end{figure}

\end{document}